\documentclass[pr,twocolumn,preprintnumbers ,amsmath, amssymb, superscriptaddress, aps]{revtex4-1}
\usepackage{color}
\usepackage{amsmath}
\usepackage{amssymb}  
\usepackage{bbold}
\usepackage{float}
\usepackage{tikz}
\usepackage{makecell}
\usepackage{subfigure}
\usepackage{graphicx} 
\usepackage{dcolumn}  
\usepackage{bm}       
\usepackage{multirow} 
\usepackage{hyperref}
\usepackage{placeins}
\newcommand{\vect}[1]{\boldsymbol{#1}}

\begin{document}

\renewcommand\floatpagefraction{0.8} 
\renewcommand\topfraction{0.8}       

\author{Martin Rodriguez-Vega}
\email{rodriguezvega@utexas.edu}
\affiliation{Department of Physics, The University of Texas at Austin, Austin, TX 78712, USA}
\affiliation{Department of Physics, Northeastern University, Boston, MA 02115, USA}
\author{Michael Vogl}
\email{ssss133@googlemail.com}
\affiliation{Department of Physics, The University of Texas at Austin, Austin, TX 78712, USA}
\affiliation{Department of Physics, King Fahd University of Petroleum and Minerals, 31261 Dhahran, Saudi Arabia}
\author{Gregory A. Fiete}
\affiliation{Department of Physics, Northeastern University, Boston, MA 02115, USA}
\affiliation{Department of Physics, Massachusetts Institute of Technology, Cambridge, MA 02139, USA}
\title{Floquet engineering of twisted double bilayer graphene}
\date{\today}

\begin{abstract}
Motivated by the recent experimental realization of twisted double bilayer graphene (TDBG) samples we study, both analytically and numerically, the effects of circularly polarized light propagating in free space and confined into a waveguide on the band structure and topological properties of these systems. These two complementary Floquet protocols allow to selectively tune different parameters of the system by varying the intensity and the light frequency. For the drive protocol in free space, in the high-frequency regime, we find that in TDBG with AB/BA stacking, we can selectively close the zone-center quasienergy gaps around one valley while increasing the gaps near the opposite valley by tuning the parameters of the drive. In TDBG with AB/AB stacking, a similar effect can be obtained upon the application of a perpendicular static electric field.  Furthermore, we study the topological properties of the driven system in different settings, provide accurate effective Floquet Hamiltonians, and show that relatively strong drives can generate flat bands. On the other hand, longitudinal light confined into a waveguide couples to the components of the interlayer hopping that are perpendicular to the TDBG sheet, allowing for selective engineering of the bandwidth of Floquet zone center quasienergy bands without breaking the symmetries of the static system. 
\end{abstract}
\maketitle

%
%
\section{Introduction}

Moir\'e superlattices have emerged as platforms to attain strongly correlated phases of matter by controlling the stacking configuration between the layers~\cite{Bistritzer12233,dosSantos2012,morell2010}. In twisted bilayer graphene (TBG) samples, examples include superconducting, Mott-insulating ~\cite{Cao2018,Cao2018sc,Codecidoeaaw9770,wong2019cascade,Lu2019efetov}, and ferromagnetic states~\cite{Sharpe605, Seo_2019}. In twisted transition metal dichalcogenide heterostructures (TMDs), evidence for moiré excitons has been reported~\cite{Tran2019,Seyler2019,Alexeev2019}. More recently, twisted double bilayer graphene (TDBG) has emerged as a multi-flat-band system, exhibiting pin-polarized and correlated phases~\cite{Cao2020,PhysRevB.99.235417,PhysRevB.99.235406,Lee2019,doi:10.1021/acs.nanolett.9b05117, C9NR10830K,PhysRevB.99.075127}. 

The plethora of strongly-correlated phases available in moir\'e superlattices naturally invites for the development of controllable mechanisms that would allow one to tune in and out of these phases. In equilibrium, hydrostatic pressure has been used to increase the tunneling strength and tune the magic angle in TBG~\cite{Yankowitz1059,PhysRevB.98.085144,Chittari_2018,Yankowitz2018,PhysRevLett.118.147401,Yue1701696}. On the other hand, out-of-equilibrium approaches, such as Floquet engineering~\cite{Oka_2009,Eckardt_2015,Blanes2009,Feldm1984,Magnus1954,PhysRevB.95.014112,Bukov_2015,PhysRevA.68.013820,PhysRevX.4.031027,PhysRevLett.115.075301,PhysRevB.93.144307,PhysRevB.94.235419,PhysRevLett.116.125301,PhysRevB.25.6622,PhysRevX.9.021037,Vogl_2019,PhysRevB.101.024303,Rodriguez_Vega_2018,Martiskainen2015,rigolin2008,weinberg2015,verdeny2013,McIver2020,PhysRevB.100.125302,PhysRevB.95.201411,PhysRevLett.121.036402,PhysRevB.98.041113,PhysRevB.100.085138,PhysRevLett.120.106601}, provide a more flexible and controllable route. Recently, the use of lasers at various frequencies have been proposed to engineer the Floquet band structure of graphene-based moir\'e superlattices. In the high frequency regime, it has been shown that topological transitions can be induced in large twist angle TBG~\cite{PhysRevResearch.1.023031} and topological flat bands with non-zero Chern numbers can be induced in the ultraviolet regime~\cite{li2019floquetengineered}. In the near-infrared range, several flat bands can be generated~\cite{katz2019floquet}. In the low-frequency regime, Floquet drives can generate a large variety of broken symmetry phases as revealed by effective Floquet Hamiltonians~\cite{vogl2020effective}. Finally, light confined into a waveguide provides a way to selectively increase or decrease the the magic angle by driving in the low- or high-frequency regime~\cite{vogl2020tuning}.

\begin{figure}[t]
	\begin{center}
		\subfigure{\includegraphics[width=8.50cm]{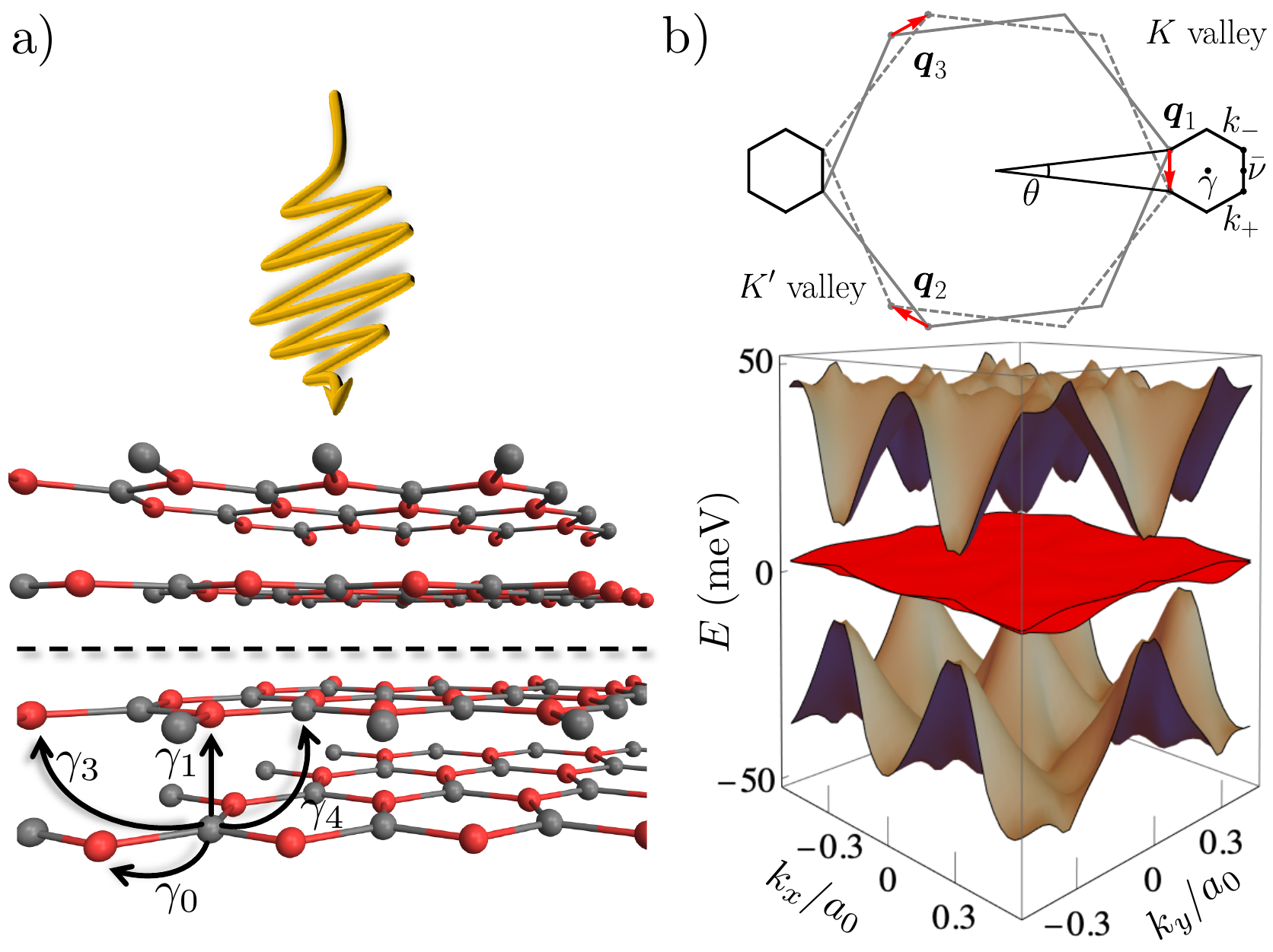}}		
		\caption{(Color online) a) Sketch of double bilayer graphene irradiated by circularly polarized light. The black arrows indicate the various tunneling process within each bilayer. The dashed black line represents the twist. b)  Moir\'e Brilloiun zone and band structure for twisted double bilayer graphene near the K point neglecting $\gamma_{3/4}$. The parameters are $w_0=100$~meV, $w_1 = 120 $~meV, and $\theta=1.05^\circ$.}
		\label{fig:fig1}
	\end{center}
\end{figure}

Floquet engineering has also been proposed for the generation of valley polarized currents in graphene, TMDs and van der Waals heterostructures~\cite{abergel2009, fengua2014, PhysRevLett.116.016802, jimnezgaln2019lightwave,Langer2018,Shin2018} with applications in valleytronics~\cite{Schaibley2016, doi:10.1002/smll.201801483}. An interesting effect with topological origin is the valley Hall effect\cite{PhysRevLett.99.236809,PhysRevB.77.235406,PhysRevLett.108.196802,PhysRevLett.110.066803}, which has been experimentally observed in monolayer TMDs illuminated with circularly polarized light~\cite{Mak1489} and graphene-hexagonal boron nitride heterostructures~\cite{Gorbachev448}. Also, in TMDs, exciton level selective tuning using intense circularly polarized light has been demonstrated~\cite{Sie2015}, and the valley Bloch-Siegert shift has been observed~\cite{Sie1066}. Furthermore, in bilayer graphene in the presence of a perpendicular electric field, valley topological transport has been reported~\cite{Sui2015, Shimazaki2015}. New flexible and controllable platforms for the manipulation of the valley degree of freedom are highly desirable for information processing. 

In this work, we consider TDBG in the AB/AB and AB/BA configurations irradiated by circularly polarized light in free-space and confined into a waveguide. We show by deriving effective Floquet Hamiltonians and by numerical calculations that light in free-space can induce transitions from a trivial or valley Chern insulator (depending on the stacking configuration) into a Chern insulator. Furthermore, in the presence of a transverse electric field,  driven AB/AB TDBG allows the quasienergy gaps at the $K$ and $K'$ valleys to be selectively tuned by varying the direction of the static electric field, and by modulating the amplitude and frequency of the driving laser. For AB/BA TDBG, we find that the quasienergy gaps can be tuned selectively even without an applied electric field. The flexibility of the quasienergy band structure near the Floquet zone center can be used to generate valley polarized currents in TDBG, independent of the stacking configuration. On the other hand, using light confined into a waveguide allows to dynamically tune the component of the tunneling perpendicular to the plane without breaking the symmetries of the static system. 

The rest of the paper is organized as follows. In Sec. \ref{sec:static} we describe static TDBG and the notation we adopt throughout the paper. In Sec. \ref{sec:driv_free}, we consider TDBG driven by circularly polarized light in free space. We consider both high- and intermediate-frequency regimes and describe effects on the band structure and the topological aspects in each regime.   In Sec. \ref{sec:waveguide}, we consider a drive protocol using longitudinal vector potentials, allowed inside a waveguide and discuss the effects on the quasienergies.  Finally, in Secs. \ref{sec:exp} and \ref{sec:conclusion}  we comment on the experimental drive parameters necessary to observe the effects before discussed and present our conclusions, respectively.

\section{Static system} 
\label{sec:static}
\begin{figure}[t]
	\begin{center}
		\subfigure{\includegraphics[width=7.50cm]{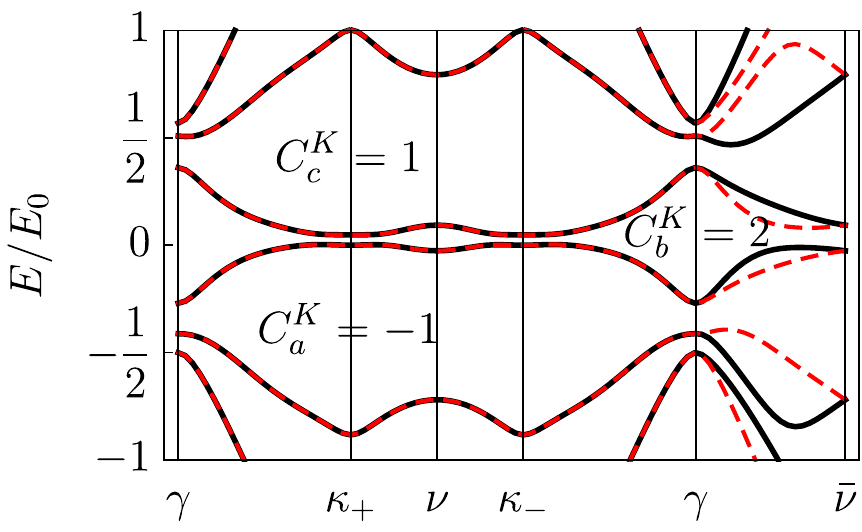}}		
		\caption{(Color online) AB/BA TDBG band structure for $\theta = 1.4^\circ$, $\Delta=0$ and $\gamma_{3/4} = 0$ along a high symmetry path in the mBZ. The black solid (red dashed) lines correspond to the spectrum near the $K$ ($K'$) point. The Chern numbers in the gaps labeled $a$, $b$, and $c$ are indicated for the $K$ point. Time reversal symmetry imposes $C^{K'}=-C^{K}$. The energy scale is $E_0=100$~meV.}
		\label{fig:fig2}
	\end{center}
\end{figure}

In the continuum limit, the static Hamiltonian for TDBG near the $K$ point with AB/AB ($s=s'=1$) [AB/BA ($s=-s'=1$)] stacking patterns is given by \cite{PhysRevB.99.235417,PhysRevB.99.235406,Lee2019}
\begin{align}\nonumber
H_{ss'}(\vect k,\vect x)  & = \tau_u \otimes h_s(-\theta/2, \vect k-\kappa_-) \\ \nonumber &  +  \tau_d \otimes h_{s'}(\theta/2,\vect k-\kappa_+) \\
&+ \tau^+ \otimes \lambda^- \otimes T(\vect x)  + \tau^- \otimes \lambda^+ \otimes T^{\dagger}(\vect x) ,
\label{twist_bilayer_Ham}
\end{align}
where $\tau_u  = \left(\mathbb 1 + \tau_3 \right)/2$, $\tau_d  = \left(\mathbb 1 - \tau_3 \right)/2$, $\tau_{\pm}  = \left(\tau_1 \pm i \tau_2 \right)/2$, and $\tau_i$ and $\lambda_i$ are Pauli matrices in top/bottom bilayer and layer space, respectively. Here, $\sigma_k$ are Pauli matrices or identity operators in pseudospin space. The bilayer graphene Hamiltonian is given by~\cite{jung2014}

\begin{align}
  h_s(\theta,\vect k)&= \left(\begin{array}{cc|cc}
       \Delta_1 + \delta^-_s & \gamma_0 f(R_\theta \vect k)    &     \multicolumn{2}{c}{\smash{\raisebox{-.5\normalbaselineskip}{$t_s(\vect k)$}}} \\
       \gamma_0 f^*(R_\theta\vect k)&\Delta_1+ \delta^+_s &           &   \\
      \hline \\[-\normalbaselineskip]
     \multicolumn{2}{c|}{\smash{\raisebox{-.5\normalbaselineskip}{$t^{\dagger}_s(\vect k)$}}} &     \Delta_2 + \delta^+_s& \gamma_0 f(R_\theta\vect k)\\
       &        &  \gamma_0 f^*(R_\theta\vect k) & \Delta_2+ \delta^-_s
    \end{array}\right),
    \label{eq:bilayer}
\end{align}
with tunneling matrix 
\begin{align}
	&t_{+}(\vect k)=
	 \begin{pmatrix}
	 - \gamma_4 f(R_\theta \vect k) & - \gamma_3 f^*(R_\theta\vect k) \\
	 \gamma_1 & - \gamma_4 f(R_\theta \vect k)
	\end{pmatrix}.
\end{align}

Each diagonal block in Eq. \eqref{eq:bilayer} corresponds to the top and bottom layers of each bilayer unit, $f(\vect k) = k_x - i k_y$ describes the intralayer hopping between nearest-neighbor sites, and $\gamma_0=v_F/a_0$ in natural units ($\hbar = c = e = 1$). Here, $\Delta_i$ corresponds to a potential on graphene layer $i$, which will describe the effect of an applied electric field perpendicular to the sample surface. Finally, $\delta^{\pm}_s = \delta( 1\pm s)/2$ is a stacking and layer-dependent gap~\cite{jung2014}.

The off-diagonal blocks $t_{s}(\vect k)$ describe the tunneling processes within each bilayer unit~\cite{jung2014}, including contributions from vertical tunneling $\gamma_1$, and next-nearest neighbor tunneling $\gamma_3$, and $\gamma_4$. $\gamma_3$ leads to trigonal warping and $\gamma_4$ to particle-hole symmetry breaking. The tunneling sector also depends on the bilayer stacking configuration $s$. 

The interlayer hopping matrix
 \begin{align}\label{eq:tunn_1}
 	&T(\vect x)=\sum_{i=-1}^1 e^{-i\vect Q_i\vect x} T_i,\\
 	&T_i=w_0\mathbb{1}_2+w_1\left(\cos\left(\frac{2\pi i}{3}\right)\sigma_1+\sin\left(\frac{2\pi i}{3}\right)\sigma_2\right),
 	\label{eq:tunn_2}
\end{align}
describes tunneling between the two graphene bilayers, where $\vect Q_0=(0,0)$, and $\vect Q_{\pm 1}= k_\theta\left(\pm \sqrt{3}/2,3/2\right)$ are the reciprocal lattice vectors. We neglect direct tunneling contributions between layers that are not adjacent to one another, as indicated by the structure $\tau^+ \otimes \lambda^- \otimes T(\vect x)$. The parameter $w_1$ in the tunneling term models relaxation effects, since the AB and BA  configurations within each bilayer units are energetically preferred over the AA configuration~\cite{PhysRevB.96.075311,fleischmann2019moir}. Throughout this work, we fix the parameters $\gamma_0=v_F/a_0=2.36$ eV, $a_0 = 2.46\mbox{ \normalfont\AA}$, $w_0=100$~meV, $w_1=120$~meV, $\gamma_3=283$~ meV, $\gamma_4=138$~ meV and $\delta=15$~meV unless otherwise explicitly stated.

The Hamiltonian near the $K'$ valley can be obtained by applying a time reversal operation $\mathcal T$ to the Hamiltonian at the $K$ valley~\cite{Balents2019}. Before studying the time dependent case it is worthwhile to summarize various symmetry properties of static TDBG.  In addition to time-reversal symmetry $\mathcal T$,  AB/AB TDBG possesses $C_{3z}$ rotational symmetry, and mirror symmetry $M_x: y, k_y \to -y,-k_y$ in the absence of an applied static electric field. The AB/BA TDBG possesses $C_{3z}$, mirror symmetry $M_y: x,k_x \to -x,-k_x$ (which switches the valleys), and $M_y \mathcal T$~\cite{PhysRevB.99.075127,Lee2019,PhysRevB.99.235406}. 

In addition, TDBG displays topological properties captured by the Chern number, which is defined by $C = \sum_{n\in \text{occ.}} C_n$, with band Chern number
\begin{equation}
C_n = \frac{1}{2 \pi} \int_{\text{mBZ}} F_n(\vect k)d \vect k,
\label{eq:chern}
\end{equation}
where $F_n(\vect k) =\left( \nabla \times \vect {\mathcal{A}}_n(\mathbf{k}) \right)_z$ is the Berry curvature, $\vect{\mathcal{A}}_{n}(\mathbf{k})=-i \langle u_{n}( \mathbf{k})\left|\partial_{\vect k}\right| u_{n}(\mathbf{k})\rangle$ the Berry connection, and  $| u_{n}(\mathbf{k})\rangle$ the eigenstates of $H_{s}(\vect k,\vect x)$ defined on a plane wave basis. Time reversal symmetry implies that the Chern numbers for each valley are opposite to each other for a given band $n$, $C^K_{n} = -C^{K'}_{n}$. In the absence of a potential difference, $\Delta_i = 0$, the $M_y$ symmetry of  AB/AB TDBG implies $C^{K/K'}_{n}=0$ for each band $n$~\cite{Lee2019}, since it does not interchange the valleys. For example, by explicit evaluation of Eq. \eqref{eq:chern} near $K$, we find that AB/AB TDBG with $\gamma_{3/4}=0$ has trivial Chern numbers $C^K_a =C^K_b = C^K_c = 0$ at gaps $\delta E_i$, where $i=a,b,c$ labels the gaps as shown in Fig. \ref{fig:fig2}. On the other hand, AB/BA TDBG has non-trivial Chern number $C^K_a = -1$, $C^K_b = 2$, and $C^K_c = 1$ even for $\Delta_i = 0$. At the $K'$ point, we find $C^{K'}_a = 1$, $C^{K'}_b = -2$, and $C^{K'}_c = -1$, as required by time-reversal symmetry, placing AB/BA TDBG in a Hall valley insulating phase.  

In the next section, we will study the effect of circularly polarized light on TDBG.

\section{Driven system in free space}
\label{sec:driv_free}

In this section, we consider the effect of circularly polarized light in free space incidenting normal to the TDBG surface. The time-dependent Hamiltonian near the $K$ point is given by $ H_{ss^\prime}(t)  \equiv H_{ss^\prime}(\vect k(t),\vect x) $, 
\begin{align}\nonumber
H_{ss'}(\vect k(t),\vect x) & = \tau_u \otimes h_s(-\theta/2, \vect k(t)-\kappa_-)\\ \nonumber &   +  \tau_d \otimes h_{s'}(\theta/2,\vect k(t)-\kappa_+) \\
&+ \tau^+ \otimes \lambda^-  \otimes T(\vect x)  +  \tau^-\otimes \lambda^+ \otimes T^{\dagger}(\vect x) ,
\label{eq:twist_bilayer_Ham_time}
\end{align}
and $k_x(t)  = k_x - A \cos (\Omega t)$, and $k_y(t) = k_y - A \sin (\Omega t)$. Here we used a minimal coupling procedure that is valid for not too strong couplings to the electromagnetic field\cite{PhysRevB.101.205140}. The vector potential enters in the same way near both the $K$ and $K'$ points. The inter-bilayer tunneling sector has, in principle, contributions parallel to the surface that could couple to the normally incident circularly polarized light. However, the orbital overlap decays exponentially away from sites that sit on top of each other in twisted sample~\cite{Bistritzer12233}. The time-dependent Hamiltonian \ref{eq:twist_bilayer_Ham_time} satisfies $H_{ss^\prime}(t+2\pi/\Omega) = H_{ss^\prime}(t)$. Therefore, we employ Floquet theory to write the wavefunctions as $| \psi (t) \rangle = e^{i \epsilon t} |\phi(t) \rangle$, where $|\phi(t + 2\pi/\Omega) \rangle = |\phi(t) \rangle$ are the steady states and $\epsilon$ is the quasienergies which satisfy the Floquet-Schr\"odinger equation
\begin{equation}
[H_{ss^\prime}(t)-i\partial_t] |\phi(t)\rangle = \epsilon |\phi ( t)\rangle.
\end{equation}
In the extended-space picture~\cite{Eckardt_2015,Mikami_2016,PhysRevA.7.2203},  $|\phi(t) \rangle  = \sum_n e^{i n \Omega t} |\phi_n \rangle$. An expansion of the operator $[H_{ss^\prime}(t)-i\partial_t]$ in modes $e^{i n \Omega t}$ leads to $\sum_m \left(H_{ss^\prime}^{(n-m)} + \delta_{n,m} \Omega m \right) |\phi_m\rangle = \epsilon|\phi_n\rangle$,
where $H_{ss^\prime}^{(n)} = \int_0^{2 \pi} d\tau/(2\pi) H_{ss^\prime}(\tau) e^{-i \tau n}$. In the next two subsections, we consider the effects of the drives in the high- and intermediate-frequency regimes, respectively.

\subsection{High-frequency}

\begin{figure}[b]
	\begin{center}
		\subfigure{\includegraphics[width=8.50cm]{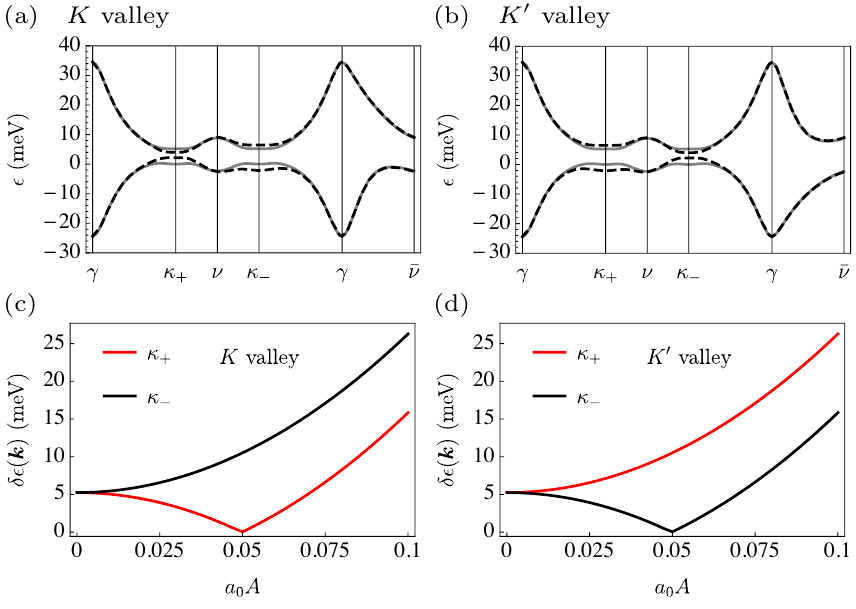}}		
		\caption{(Color online) (a)((b)) AB/AB TDBG quasienergies near the $K$($K'$) point along a high-symmetry path in the mBZ for $\theta = 1.4^\circ$, $\Omega/W=2$, and $a_0 A=0.04$. The gray curves correspond the equilibrium energies. (c)((d)) Quasienergy gap for at the $\kappa_{\pm}$ as a function of the driving strength $a_0 A$ for $\Omega/W=2$. At the $K$ valley, the $\kappa_+$ closes at $a_0 A \approx 0.05$. At the $K'$ valley, $\kappa_-$ closes at the same driving strength.}
		\label{fig:gapabab}
	\end{center}
\end{figure}

In the high-frequency regime, we employ the van Vleck expansion~\cite{Eckardt2015b} to obtain an effective Floquet Hamiltonian $H^s_{\text{Vv}} = H_s^{(0)} + \delta H_{s,\text{vV}}$, where $H_s^{(0)}$ is the Hamiltonian averaged over one drive period, and  
\begin{align}\nonumber
\delta H_{ss',\text{vV}} &= -(\Delta_{\text{vV}}-\Delta^{(3)}_{\text{vV}}) \mathbb{1} \otimes \mathbb{1} \otimes  \sigma_3 - (\Delta^{(4)}_{\text{vV}} - \Delta^{(3)}_{\text{vV}}) \times \\ 
& \left( s \tau^u  \otimes  \lambda_3  \otimes \mathbb{1} + s' \tau^d \otimes  \lambda_3  \otimes \mathbb{1} \right)  
\label{eq:HvV}
\end{align}
where $\Delta_{\text{vV}}= \xi (v_F A)^2/\Omega$,  $\Delta^{(4)}_{\text{vV}}=\xi (v_4 A)^2/\Omega$, and $\Delta^{(3)}_{\text{vV}}=\xi (v_3 A)^2/(2\Omega)$, where $\xi=1$($\xi=-1$) near the $K$($K'$) valley. The simplicity of the high-frequency regime van Vleck expansion allows us to also retain the effects of $\gamma_{3,4}$, which are harder to capture using more sophisticated intermediate frequency regime methods introduced in later sections.

The gap $\Delta_{\text{vV}}$ is generated due to the effect that light has on the hopping in each graphene layer, which is captured by the component $\mathbb{1}\otimes\mathbb{1}\otimes[\sigma\cdot(\vect k-\vect A(t))]$ in the time dependent Hamiltonian. It breaks both time-reversal $\mathcal T$ and mirror symmetries ($M_x$ for AB/AB and $M_y$ for AB/BA) and is staking-independent. The gaps $\Delta^{(3,4)}_{\text{vV}}$ are induced by the effect that light has on interlayer hoppings that have components in the plane. Specifically, these interlayer hoppings are within the top and bottom bilayers and are captured by the terms $(\tau^u \otimes \lambda^+ \otimes  t_s(\vect p-\vect A(t)) + \tau^d \otimes \lambda^+ \otimes  t_{s'}(\vect p-\vect A(t)) ) + h.c.$ in the time-dependent Hamiltonian. Specifically,  $\Delta^{(4)}_{\text{vV}}$ is induced by hopping between equivalent lattice sites on opposite layers. It constitutes a potential difference between the graphene layers in each bilayer unit, and breaks both time-reversal and mirror symmetries in both stacking configurations. On the other hand, $\Delta^{(3)}_{\text{vV}}$  is caused by hoppings between inequivalent sublattices on opposite layers. It has two components. The first one is independent of the stacking configuration and acts as $\Delta_{\text{vV}}$. The second component depends on the stacking configuration (AB/AB or AB/BA), and acts as $\Delta^{(4)}_{\text{vV}}$.  Next, we will review the effect of these dynamically-induced terms on the topological properties of TDBG.

\begin{figure}
	\begin{center}
		\subfigure{\includegraphics[width=8.50cm]{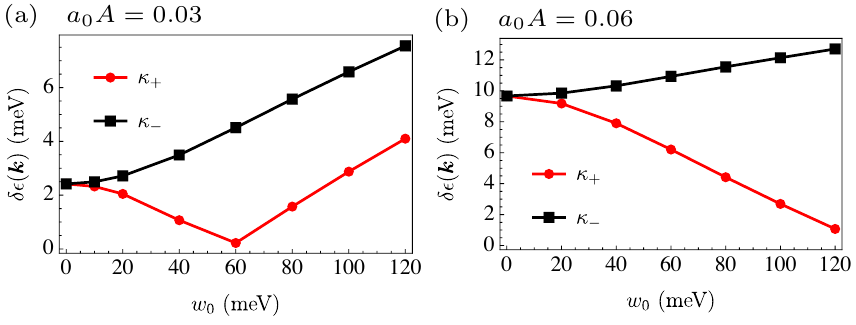}}		
		\caption{(Color online) (a)((b)) AB/AB TDBG quasienergy gaps at $\kappa_{\pm}$ near the $K$ valley for $\theta = 1.4^\circ$, $\Omega/W=2$, and (a)((b)) $a_0 A=0.03$ ($a_0 A=0.06$) as a function of the tunneling between the bilayer units. We set $w_0 = w_1$ in this case.}
		\label{fig:gaptunn}
	\end{center}
\end{figure}

In equilibrium,  AB/AB TDBG is a trivial insulator for $\Delta_i=0$. The finite $\Delta_{\text{vV}}$ induced by circularly-polarized light leads to a transition into a Chern insulator with Floquet topological bands. For example, consider the case $\theta =1.4^\circ$, $\gamma_{3/4}=0$, $\Omega/W=2$, with $W = v_F/a_0$. In Figs. \ref{fig:gapabab}(a-b), we show the quasienergy spectrum near the Floquet zone center $\epsilon/\Omega = 0$ at the $K$ and $K'$ valleys, respectively. The effect of light at the $\kappa_{\pm}$ near each valley is the opposite of each other. In Figs. \ref{fig:gapabab}(c-d), 
we show the evolution of the quasienergy gap at $\kappa_{\pm}$ as a function of the drive strength obtained numerically by diagonalizing the Hamiltonian in the Floquet extended space. When the gap closes at $\kappa_+$ for $a_0 A \approx 0.05$, and then opens again, the band Chern number changes from zero to $C^K_{n=-1}=-2$ for the lower Floquet zone center quasienergy band (labeled as $n=-1$) and $C^K_{n=1} =  2$ for the higher quasienergy band. This can be understood from the sum of the contributions of the Berry curvature from the four Dirac cones composing TDBG near the $K$ valley. At the $K'$ valley, we find $C^{K'}_{n} = C^K_n$, since the restrictions from time-reversal symmetry are lifted. The asymmetric behavior of the gaps at $\kappa_{\pm}$ arises because the hybridization of the twisted bilayers breaks inversion symmetry, and there is no $C_{2z}$ rotational symmetry as in TDG, leading to a generic gapped state in the absence of a drive. Upon the application of the drive, the states at $\kappa_{\pm}$ evolve in time in distinct manners resulting in the structure of $\Delta_{\text{vV}}$ in the effective Floquet Hamiltonian. In Fig. \ref{fig:gaptunn}, we show the evolution of the gap at $\kappa_{\pm}$  as a function of the tunneling amplitude between the twisted bilayers for drive strengths $a_0 A=0.03$ and $a_0 A=0.06$ (below and above the light-induce transition for the nominal values $w_0=100$~meV and $w_1=120$~meV). For fully decoupled layers ($w_0=w_1 = 0$), the gaps are symmetric. 

After discussing the Chern number, we recall that while it is a measurable quantity \cite{PhysRevA.91.043625,PhysRevB.101.174314}, it is not what determines the number of edge states. Rather, in Floquet systems the bulk-edge correspondence is determined by the winding number $\mathcal W[U_{\varepsilon}]$, defined at a quasienergy $\varepsilon$ inside a gap~\cite{PhysRevX.3.031005}, where 
\begin{align}
\mathcal W[\mathcal U]=& \frac{1}{8 \pi^{2}} \int dt d \vect k 
  \text{Tr}\left(\mathcal U^{-1} \partial_{t} \mathcal U\left[\mathcal U^{-1} \partial_{k_{x}} \mathcal U, \mathcal U^{-1} \partial_{k_{y}} \mathcal U\right] \right),
\end{align}
and $\mathcal U_{\varepsilon}$ is a modified time evolution operator~\cite{PhysRevX.3.031005}. Here, we calculate $\mathcal W$  via the truncated Floquet Hamiltonian in the extended space~\cite{PhysRevX.3.031005}. For the  AB/AB TDBG case above, we find $\mathcal W^K_a=0$, $\mathcal W^K_b=-2$,  $\mathcal W^K_c=0$,and $\mathcal W^{K'} = \mathcal W^K$ at the three gaps considered around the quasienergy bands shown in Fig. \ref{fig:gapabab}(a-b).

Now let's consider the AB/BA configuration for TDBG with $\Delta_i=0$. Contrary to AB/AB TDBG, AB/BA TDBG is a valley Chern insulator at equilibrium. The energies and Chern numbers inside the gaps are shown in Fig. \ref{fig:fig2}. As for the case of AB/AB TDBG, circularly polarized light leads to a transition into a Chern insulating phase with finite Floquet band Chern and winding numbers. However, in this configuration, the behavior of the $\kappa_{\pm}$ gaps is different: at the $K$ valley, both $\kappa_{\pm}$ gaps close at drive amplitude $a_0 A \approx 0.058$ for $\Omega/W=2$, while the $\kappa_{\pm}$ gaps near the $K'$ valley increase monotonically with $a_0 A$. This selective gap engineering could be employed to generate valley-polarized currents in AB/BA TDBG. As for the topological properties, the Floquet band Chern numbers switch after the transition: $C^K_{n=-2}=-1$, $C^K_{n=-1}=-1$, $C^K_{n=1}=3$, and $C^K_{n=2}=-1$. The  winding numbers inside the gaps are $\mathcal W^{K}_a=-1$, $\mathcal W^{K}_b=-2$, and $\mathcal W^{K}_c=1$, with a change from gap to gap in correspondence with the Floquet band Chern numbers.

Since the gaps do not close at the $K'$ point in the range of parameters we considered, the band Chern numbers remain the same as in the static case: $C^{K'}_{n=-2}=1$, $C^{K'}_{n=-1}=-3$, $C^{K'}_{n=1}=1$, and $C^{K'}_{n=2}=1$. For the winding numbers, we obtain $W^{K'}_a=1$, $W^{K'}_b=-2$, and $W^{K'}_c=-1$. 

\begin{figure}[t]
	\begin{center}
		\subfigure{\includegraphics[width=8.50cm]{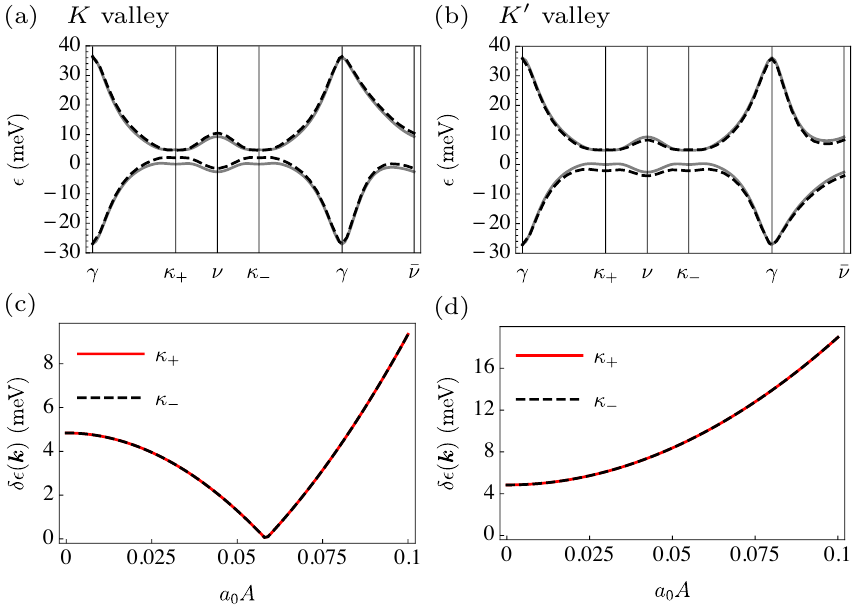}}		
		\caption{(Color online) (a)((b)) AB/BA TDBG quasienergies near the $K$($K'$) point along a high-symmetry path in the mBZ. The parameters are the same as in Fig. \ref{fig:gapabab}. The gray curves correspond the equilibrium energies. (c)((d)) Quasienergy gap for at the $\kappa_{\pm}$ as a function of the driving strength $a_0 A$ for $\Omega/W=2$.}
		\label{fig:gapabba}
	\end{center}
\end{figure}

\begin{figure}[t]
	\begin{center}
		\subfigure{\includegraphics[width=8.50cm]{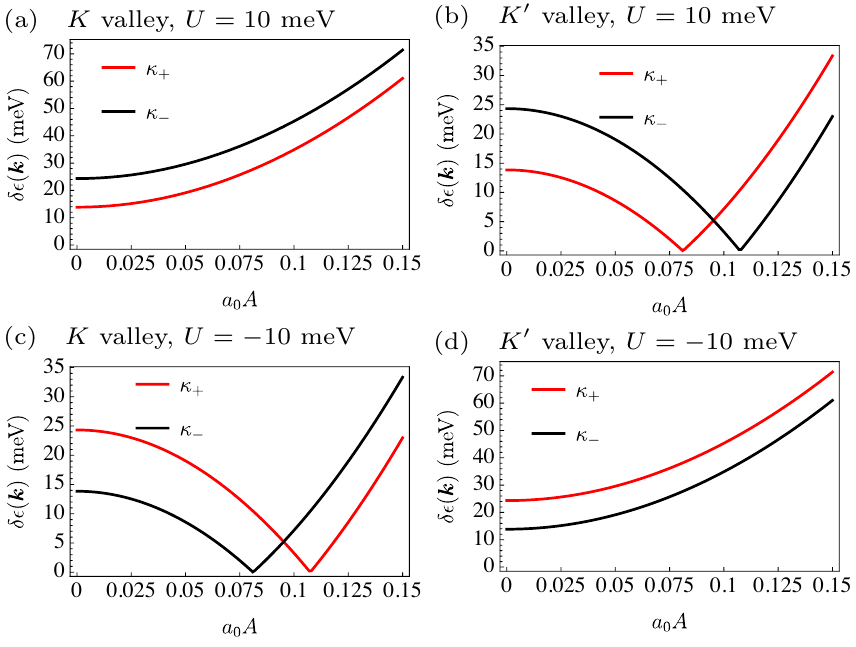}}		
		\caption{(Color online) (a)((b)) AB/AB TDBG quasienergy gap near the $K$($K'$) valley at  $\kappa_{\pm}$ for $U=10 $~meV. Panels (c-d) show the results for $U=-10 $~meV. The rest of the parameters are the same as in Fig. \ref{fig:gapabab}. }
		\label{fig:gapababU}
	\end{center}
\end{figure}

\textit{Applied static electric field}. So far, we have restricted our analysis to $\Delta_i=0$, which corresponds to no potential difference between the layers (apart from $\delta^{\pm}_s$). Now we consider AB/AB TDBG in the presence of an applied perpendicular static electric field, which leads to a potential differences between the layers. We set $\Delta_4=-\Delta_1$, $\Delta_3=-\Delta_2$, $\Delta_1=3 U/2$, and $\Delta_2= U/2$, with $U$ being the potential difference. In equilibrium, the transverse electric field places AB/AB TDBG in a valley Chern insulating regime. For $U=10$~meV, we find in the static case the band Chern numbers $C^K_{-1}=-2$ and $C^K_{1}=2$ and corresponding total Chern numbers inside the gaps  $C^K_{a}=0$, $C^K_{b}=-2$, and $C^K_{c}=0$. At the $K'$ valley, we find $C^{K'}_{n}=-C^{K}_{n}$, as imposed by time-reversal symmetry. When one drives the system, the gaps at the Floquet zone center are renormalized. In Fig. \ref{fig:gapababU}(a), we plot the evolution of the quasienergy gaps at the $\kappa_{\pm}$ points as a function of the drive amplitude. Near the $K$ valley, the quasienergy differences at $\kappa_{\pm}$ increase monotonically with $a_0 A$. Since the gap remains open for the driving parameters considered, the Chern and winding numbers do not change. In contrast, at the $K'$ point, the quasienergy differences decrease starting from different values in the vanishing drive strength limit, leading to a gap closing at $\kappa_{+}$ for $a_0 A \approx 0.08$, followed by a closing at $\kappa_{-}$ for $a_0 A \approx 0.107$. At the first quasienergy gap closing, the winding number changes from $\mathcal W^{K'}_{b}=2$ to $\mathcal W^{K'}_{b}=0$, and after the second gap closing to $\mathcal W^{K'}_{b}=-2$. The gap behavior at the $K$ and $K'$ valleys can be switched by changing the sign of the applied electric field, as shown in Fig. \ref{fig:gapababU}(c-d).


\textit{Trigonal warping and particle-hole symmetry-breaking terms.} The structure of the van Vleck Hamitonian \eqref{eq:HvV} shows that trigonal warping ($\gamma_3$ term in the bilayer graphene tunneling sector) induces a small correction to the gap $\Delta_{\text{vV}}$, since $\Delta^{(3)}_{\text{vV}}/\Delta_{\text{vV}} \approx  0.014$, independent of the frequency and amplitude of the drive. However, the effect in the static energies is not negligible. 

\begin{figure}[t]
	\begin{center}
		\subfigure{\includegraphics[width=8.50cm]{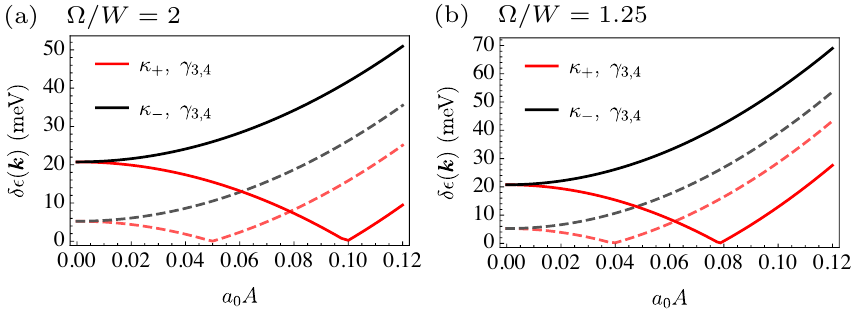}}		
		\caption{(Color online) AB/AB TDBG quasienergy gap near the $K$ valley at  $\kappa_{\pm}$ $\gamma_{3,4}=0$ (dashed curves) and $\gamma_{3,4}\neq 0$ (solid lines) as a function of the driving strength for (a) $\Omega/W=2$, and $\Omega/W=1.25$. }
		\label{fig:warp}
	\end{center}
\end{figure}

Combined particle-hole asymmetry ($\gamma_4$) and trigonal warping effects induce a staking-dependent gap. For AB/AB stacking, $s=s'=1$, the gap has the structure  $(\Delta^{(4)}_{\text{vV}} - \Delta^{(3)}_{\text{vV}}) \mathbb{1}  \otimes  \lambda_3  \otimes \mathbb{1} $, which constitutes a potential difference between the graphene composing each bilayer, with $|(\Delta^{(4)}_{\text{vV}} - \Delta^{(3)}_{\text{vV}})/\Delta_{\text{vV}} |\approx 0.01$. For AB/BA stacking, $s=-s'=1$, the gap has the form $(\Delta^{(4)}_{\text{vV}} - \Delta^{(3)}_{\text{vV}})    \tau_3  \otimes  \lambda_3  \otimes \mathbb{1} $. In general, this terms can renormalize the topological transition points. For example, in Fig. \ref{fig:warp}, we plot the quasienergy gap at the $\kappa_{\pm}$ points as a function of the driving strength $a_0 A$ for two frequencies in the high-frequency regime. Therefore, although the static energies can be significantly modified by $\gamma_{3,4}\neq 0$, Floquet drives can be used to manipulate the gap structure. 

In this subsection, we restricted the discussion to high-frequency and weak drives. In the next subsection, we will derive an effective Floquet Hamiltonian valid for intermediate frequencies and intermediate drive strengths. We will show, in particular, that in this regime we can generate Floquet flat bands, which are impaired by trigonal warping and particle-hole symmetry-breaking effects.

\subsection{Intermediate-frequency}

In the intermediate-frequency and intermediate drive strength regime, we obtain an effective Hamiltonian by performing a modified rotating frame transformation~\cite{vogl2020effective} and taking an average over one period (see Appendix \ref{app:effective} for details). For our analytical results, we neglect the effect of next-to-nearest neighbor hopping within each bilayer unit ($\gamma_3 = \gamma_4 = 0$), but we will discuss them numerically. Then, the effective Floquet Hamiltonian is given by
\begin{align}
H^{ss'}_{F} = R^{\dagger} \left( \bar H_{ss'} + \delta H_{F} \right) R,
\end{align}
where $R$ is a twist-angle dependent unitary transformation (see Appendix \ref{app:effective} for the explicit expression), and $ \delta H_{F} =  \Delta_F  \mathbb{1} \otimes \mathbb{1} \otimes \sigma_3 $, with $\Delta_F = A J_1(2 \sqrt{2} A/\Omega)/\sqrt{2}$, where $J_n(z)$ correspond to the $n$-th Bessel function of the first kind. As in the high-frequency regime, $\delta H_{F}$ is independent of the AB/AB or AB/BA stacking configuration. $\bar H_{ss^\prime}$ is given by 
\begin{align}\nonumber
\bar{H}_{ss'}(\vect k,\vect x) & = \tau_u \otimes \tilde h_s(-\theta/2, \vect k-\kappa_-) \\ \nonumber &  +  \tau_d \otimes \tilde h_{s^\prime}(\theta/2,\vect k-\kappa_+) \\
&+ \tau^+ \otimes \lambda^- \otimes \tilde T(\vect x)  + \tau^- \otimes \lambda^+ \tilde \otimes T^{\dagger}(\vect x),
\label{twist_bilayer_Ham2}
\end{align}
where 
\begin{figure}[b]
	\begin{center}
		\subfigure{\includegraphics[width=8.50cm]{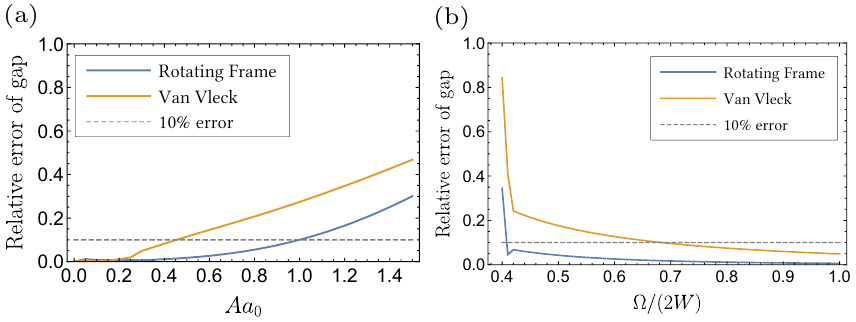}}		
		\caption{(Color online) Relative error of the quasienergy gap at the $\kappa_+$  point in the mBZ for $\delta=0$ meV as a function of (a) the driving strength for $\Omega/W = 2 $ and (b) the frequency for $a_0 A = 0.3$. In the whole range considered, the rotating frame effective Hamiltonian provides a more accurate approximation to the exact gap.}
		\label{fig:error}
	\end{center}
\end{figure}
\begin{align}
  \tilde h_s(\theta,\vect k)&= \left(\begin{array}{cc|cc}
       \Delta_1 + \tilde \delta^-_s & \tilde \gamma_0 f(R_\theta \vect k)    &     \multicolumn{2}{c}{\smash{\raisebox{-.5\normalbaselineskip}{$\tilde t_s$}}} \\
      \tilde  \gamma_0 f^*(R_\theta\vect k)&\Delta_1+ \tilde \delta^+_s &           &   \\
      \hline \\[-\normalbaselineskip]
     \multicolumn{2}{c|}{\smash{\raisebox{-.5\normalbaselineskip}{$\tilde t^{\dagger}_s $}}} &     \Delta_2 + \tilde \delta^+_s&\tilde  \gamma_0 f(R_\theta\vect k)\\
       &        & \tilde  \gamma_0 f^*(R_\theta\vect k) & \Delta_2+ \tilde \delta^-_s
    \end{array}\right),
    \label{eq:bilayer_rot}
\end{align}
$\tilde \gamma_0 = J_0(2 A/\Omega) \gamma_0 = J_0(2 A/\Omega) v_F/a_0 $, which is interpreted as a reduction of the Fermi velocity. The layer and stacking dependent gap  $\tilde \delta^{\pm}_s = \delta J_0(2\sqrt{2} A/\Omega) ( 1\pm s)/2$ is suppressed, and the tunneling is now given by $\tilde t_s= \gamma_1 J_0(2 A/\Omega) (\sigma_1-i s\sigma_2) /2$. None of these effects are captured in a leading-order van Vleck expansion, and its challenging to capture the functional form simply by computing higher-order terms. 
\begin{figure}[t]
	\begin{center}
		\subfigure{\includegraphics[width=8.50cm]{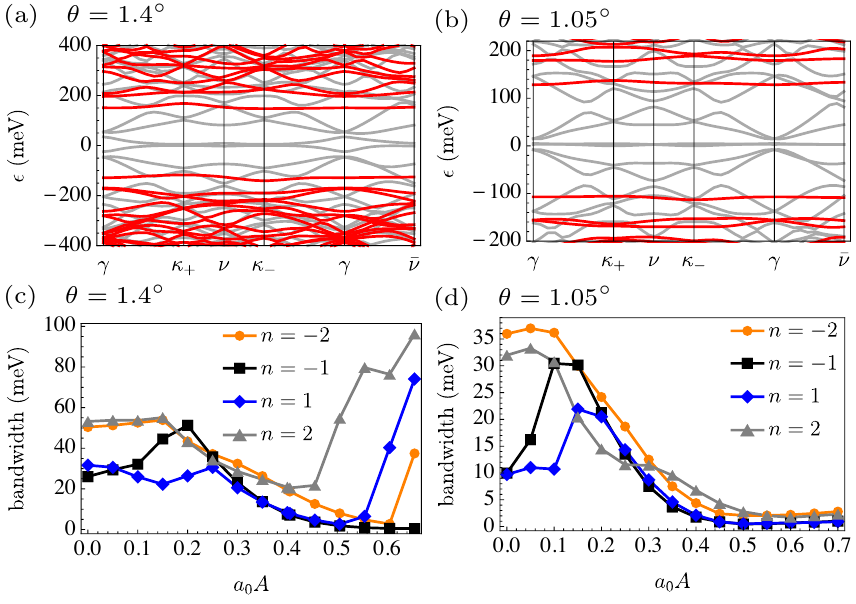}}		
		\caption{(Color online) (a) AB/AB TDBG quasienergies near the $K$ valley along a high-symmetry path in the mBZ for $\theta = 1.4^\circ$, $\Delta_i=0$, $\Omega/W=1.05$, and $a_0 A=0.3$. The gray curves correspond the equilibrium energies. In panel (b), we show the case $\theta = 1.05^\circ$ for the same drive parameters. (c-d) Bandwidth of the Floquet bands nearest ($|n|=1$) and next to nearest ($|n|=2$) to the Floquet zone center for $\theta = 1.4^\circ$ and $\theta = 1.05^\circ$, respectively, as a function of the drive strength.}
		\label{fig:qe_inter}
	\end{center}
\end{figure}
The position dependent interlayer coupling for the two center graphene layers renormalizes to
\begin{equation}
\begin{aligned}
&\tilde T(\vect x)=\sum_{n=-1}^1 e^{-i \vect Q_n \cdot \vect x}(\tilde T_n-i\omega_\theta\sigma_3)\\
&\tilde T_n=\tilde \omega_0 \mathbb{1}_2+\tilde \omega_1 \left[\cos\left(\frac{2\pi n}{3}\right)\sigma_1+\sin\left(\frac{2\pi n}{3}\right)\sigma_2\right]
\end{aligned},
\end{equation}
where
\begin{equation}
	\begin{aligned}
	&\tilde \omega_1=J_0(2 A/\Omega)\omega_1\\
	&\tilde \omega_0=\omega_0+\sin ^2(\theta/2 ) \left(J_0\left(\frac{2 \sqrt{2} A}{\Omega }\right)-1\right)\omega_0
	\end{aligned}
\end{equation} are renormalized interlayer couplings and a new angle-dependent coupling
\begin{equation}
\omega_\theta=\frac{1}{2} \sin ( \theta ) \left(J_0\left(\frac{2 \sqrt{2} A}{\Omega }\right)-1\right)\omega_0,
\end{equation} has been introduced that is absent from the equilibrium case.

As it can be deduced from Fig. \ref{fig:error}, this effective Hamiltonian is accurate up to frequency and driving strength regimes where the van Vleck approximation breaks down. In particular for a driving frequency $\Omega/W=2$ one can describe gaps with errors below $10\%$ up to driving strengths $a_0A \approx 1$, in contrast the van Vleck approximation only manages to do so until $a_0A \approx 0.45$. Therefore, the implementation of an improved transformations into a rotating frame can enhance the range of validity of effective Floquet Hamiltonians when it comes to driving strengths. A similar observation can be made if one keeps the driving strength fixed - in our case $a_0A=0.3$ - and varies the frequency. The rotating frame Hamiltonian here describes gaps with an error of less than $10\%$ for frequencies as low as $\Omega/(2W)=0.45$, while the van Vleck expansion has the same level of accuracy only up to $\Omega/(2W)=0.75 $ . Therefore, the approach allows one to reach into an intermediate strength and intermediate frequency regime, while the van Vleck expansion is restricted to large frequencies and weak coupling. This type of effective Hamiltonian could make it easier to simultaneously describe the effects of circularly polarized light for a wide range of driving protocols  and computationally challenging additional effects such as disorder.

Finally, in Fig. \ref{fig:qe_inter}(a-b), we plot the quasienergy spectrum around the Floquet zone center and along a high symmetry path in the mBZ for drive frequency $\Omega/W = 1.05$ and drive strength $a_0 A = 0.3$. We included the effects of trigonal warping and particle-hole asymmetry. In Fig. \ref{fig:qe_inter}(c-d), we show the bandwidth of the Floquet bands nearest ($|n|=1$) and next to nearest ($|n|=2$) to the Floquet zone center as a function of $a_0 A $. Therefore, stronger drives can generate Floquet flat bands, even in the presence of trigonal warping and particle-hole asymmetry, which in equilibrium tend to endow the bands with significant dispersion~\cite{Lee2019}.

\section{Driven system in a waveguide} 
\label{sec:waveguide}

\begin{figure}[h]
	\begin{center}
		\subfigure{\includegraphics[width=8.50cm]{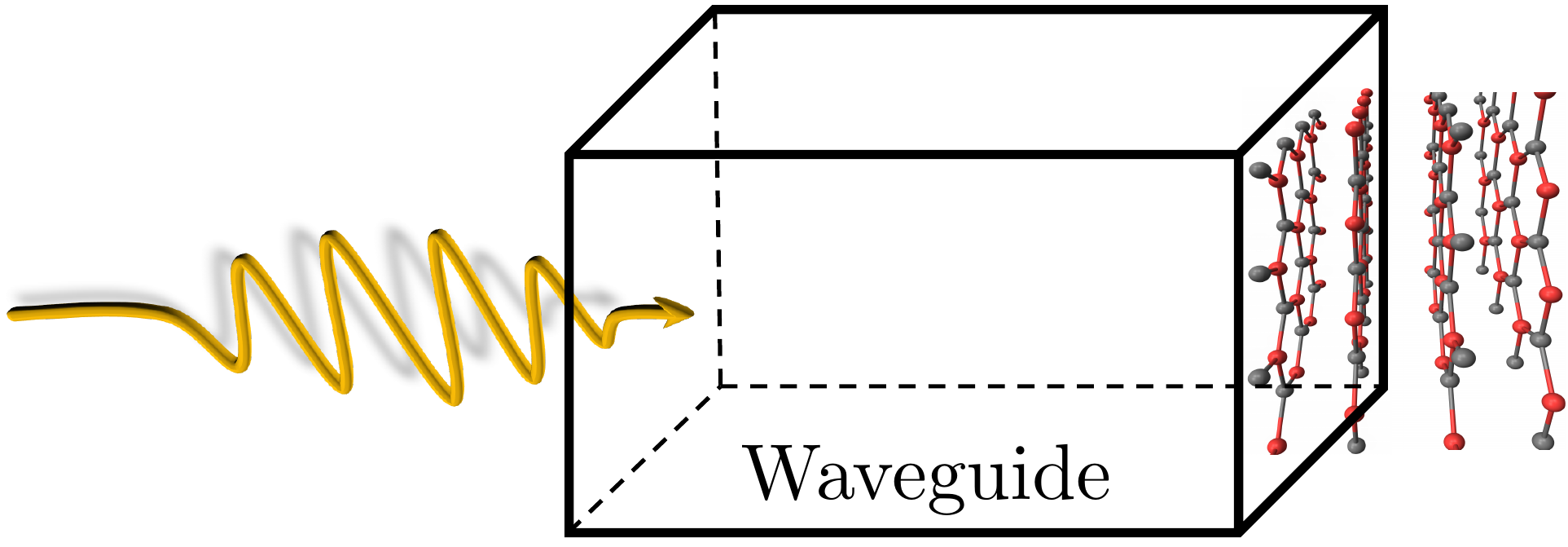}}		
		\caption{(Color online) Sketch of double bilayer graphene irradiated by circularly polarized light in a waveguide. The boundary condition imposed by the metallic walls allow for transverse modes in the electric field.}
		\label{fig:fig_wg}
	\end{center}
\end{figure}

In this section we consider a complementary Floquet protocol based on the use of light confined inside a waveguide. The boundary conditions imposed by the metallic surfaces of the waveguide allow a non-zero longitudinal component in the vector potential, which can couple to the component of the tunneling perpendicular to the TDBG plane. In the high-frequency regime, this protocol allows one to directly decrease the tunneling amplitude without breaking symmetries. 

As in free-space, the effect of the drive in a waveguide enters through a Peierls substitution to the hopping term in a real-space tight binding Hamiltonian. Particularly, the hopping term between two sites $\vect R$ and $\vect R^\prime$ acquires the position-dependent phase $c_{\vect R}^\dag c_{\vect R^\prime}\to e^{-i\int_{\vect R}^{\vect R^\prime} d\vect l \vect A}c_{\vect R}^\dag c_{\vect R^\prime}$, where $c_{\vect R^\prime}$ and $c_{\vect R}^\dag$ are the creation and a annihilation operators. Assuming that the  tight binding is defined in the $x$-$y$ plane, and the longitudinal vector potential  $\vect A(\vect r,t)\approx A(t)\hat e_z$ is incident in the $z$-direction and constant, $c_{\vect R}^\dag c_{\vect R^\prime}\to e^{-il_zA}c_{\vect R}^\dag c_{\vect R^\prime}$ with $l_z \equiv ({\vect R^\prime}-\vect R)_z$. 

For TDBG, the time-dependent Hamiltonian in the continuum limit is given by
\begin{align}\nonumber
H_{ss'}(\vect k,\vect x,t)  & = \tau_u \otimes h_s(-\theta/2, \vect k-\kappa_-,t) \\ \nonumber &  +  \tau_d \otimes h_{s'}(\theta/2,\vect k-\kappa_+,t) \\
&+ \tau^+ \otimes \lambda^- \otimes T(\vect x,t)  + \tau^- \otimes \lambda^+ \otimes T^{\dagger}(\vect x,t) ,
\label{twist_bilayer_Ham_waveguide}
\end{align}
where 
\begin{align}
  h_s(\theta,\vect k,t)&= \left(\begin{array}{cc|cc}
       \Delta_1 + \delta^-_s & \gamma_0 f(R_\theta \vect k)    &     \multicolumn{2}{c}{\smash{\raisebox{-.5\normalbaselineskip}{$t_s(\vect k,t)$}}} \\
       \gamma_0 f^*(R_\theta\vect k)&\Delta_1+ \delta^+_s &           &   \\
      \hline \\[-\normalbaselineskip]
     \multicolumn{2}{c|}{\smash{\raisebox{-.5\normalbaselineskip}{$t^{\dagger}_s(\vect k,t)$}}} &     \Delta_2 + \delta^+_s& \gamma_0 f(R_\theta\vect k)\\
       &        &  \gamma_0 f^*(R_\theta\vect k) & \Delta_2+ \delta^-_s
    \end{array}\right),
    \label{eq:bilayer_waveguide}
\end{align}
\begin{align}
	&t_{+}(\vect k,t)=
	 \begin{pmatrix}
	 - \gamma_4 f(R_\theta \vect k) & - \gamma_3 f^*(R_\theta\vect k) \\
	 \gamma_1 & - \gamma_4 f(R_\theta \vect k)
	\end{pmatrix} e^{- i a_{AB}A(t)} .
\end{align}
The tunneling sector also depends on the bilayer stacking configuration $s$. Finally, the interlayer hopping matrix acquires a time dependence according to
 \begin{align}
 	&T(\vect x,t)=\sum_{i=-1}^1 e^{-i\vect Q_i \cdot \vect x} T_i(t),\\ \nonumber
 	&T_i= e^{- i a_{AA}A(t)}w_0\mathbb{1}_2+ e^{- i a_{AB}A(t)} w_1 \times \\ &\left(\cos\left(\frac{2\pi i}{3}\right)\sigma_1+\sin\left(\frac{2\pi i}{3}\right)\sigma_2\right).
 	\label{eq:tunn_2_waveguide}
\end{align}

The above Hamiltonian is obtained by performing the substitutions
\begin{figure}
	\begin{center}
		\subfigure{\includegraphics[width=6.50cm]{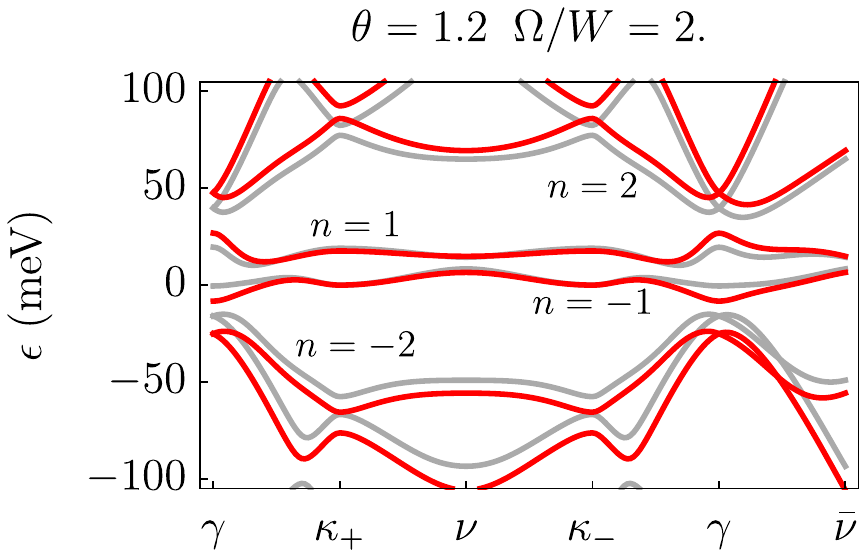}}		
		\caption{(Color online) Quasienergies along a high-symmetry path in the mBZ for AB/AB TDBG driven with light confined into a waveguide for $a_{AB} A \approx 0.136$. The gray lines correspond to the static energies. The index $n$ labels the bands closest to the Floquet zone center. }
		\label{fig:wg_qe}
	\end{center}
\end{figure}
\begin{equation}
	\begin{aligned}
	&\tau_{u,d}\otimes\lambda^\pm\otimes\sigma_i\to\tau_{u,d}\otimes\lambda^\pm\otimes\sigma_ie^{\mp i a_{AB}A(t)}\\
	&\tau^\pm\otimes\lambda^{\mp}\times\mathbb{1}\to\tau^\pm\otimes\lambda^{\mp}\times\mathbb{1}e^{\mp i a_{AA}A(t)}\\
	&\tau^\pm\otimes\lambda^{\mp}\times\sigma_{1,2}\to\tau^\pm\otimes\lambda^{\mp}\times\mathbb{1}e^{\mp i a_{AB}A(t)}\\
	\end{aligned}
	\label{interlayer_peierls}
\end{equation}
as one can confirm when mapping the corresponding tight binding hopping processes onto an effective single-particle Hamiltonian. Here, $a_{AA,AB}$ describes the interlayer distance in $AA$ and $AB$ stacked bilayer graphene. The functional form of $T(\vect x, t)$ implies that the coupling $\tau^\pm\otimes\lambda^{\mp}\times\sigma_{1,2}$ is dominant in $AB$ regions and $\tau^\pm\otimes\lambda^{\mp}\times\mathbb{1}$ is dominant in $AA$ regions. We have used this to get the approximate form in the second and third line of Eq. \eqref{interlayer_peierls}. This approximation is a simplification over the general position dependence of $T(\vect x) \to T(\vect x)e^{- i a(x,y)A}$, where the distance between layers would vary smoothly in space.

\begin{figure}
	\begin{center}
		\subfigure{\includegraphics[width=8.50cm]{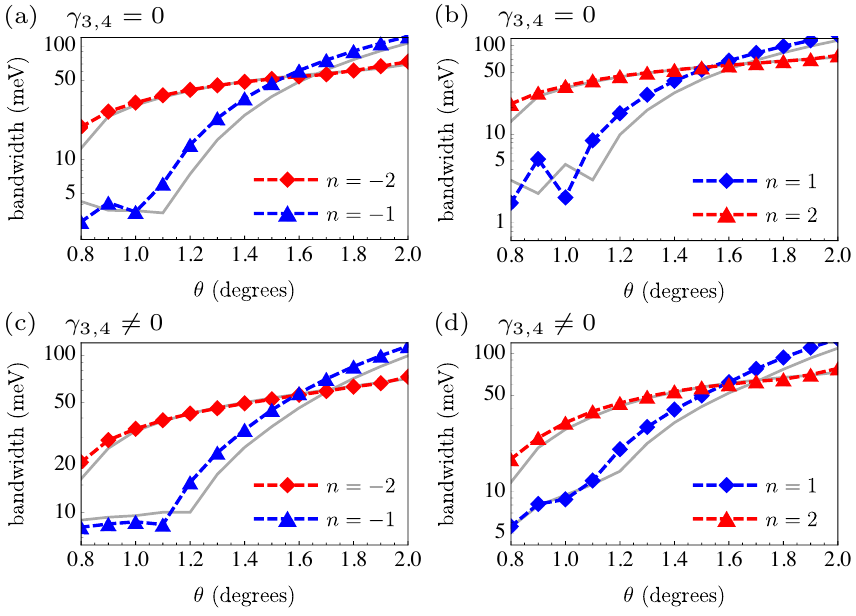}}		
		\caption{(Color online) Bandwidth of the four Floquet bands closest to the Floquet zone center as a function of the twist angle. In the top panels (a-b), we consider the case $\gamma_{3,4}=0$, and in the bottom panels (c-d), we show the case  $\gamma_3=283$~ meV, $\gamma_4=138$~ meV. The drive parameters are $\Omega/W=2$, and $a_{AB} A \approx 0.136$. The gray lines correspond to the static case. }
		\label{fig:wg_bw}
	\end{center}
\end{figure}

Specifically, we consider a transverse magnetic (TM) mode of light at the exit of a waveguide as shown in Fig. \ref{fig:fig_wg}, which for a finite region in space can have the form $\vect A\approx A\cos(\Omega t)\hat e_z$ \cite{vogl2020tuning}. In the high-frequency regime, we obtain an effect Floquet Hamiltonian using a van Vleck expansion to first order $H_{eff}\approx H_0+\sum_{m\neq 0}[H_{-m},H_m]/(2m\Omega)$, where $H_m=1/T \int_0^T H (t) e^{-im\Omega t}$. The corrections of order $1/\Omega$ vanish  if derivatives $\partial_i T(\vect x,t)$  are neglected. This is justified because all derivatives  in $H_m$ and $m\neq 0$ appear with a pre-factor $\gamma_{3,4}$ that is small and the terms $[H_{-m},H_m]/(2m\Omega)$ are already suppressed by $1/\Omega$.
In the small-angle regime, where $T(\vect x,t)$ varies slowly in real space this approximation becomes even better because then the corrections that would arise have an additional small factor $\theta$. Therefore, the leading correction is given by the averaged Hamiltonian $H_0$, which shares the same structure of the static Hamiltonian with renormalized parameters 
\begin{equation}
	\begin{aligned}
	&(w_1,\gamma_{1,3,4})\to (w_1,\gamma_{1,3,4})J_0(a_{AB}A)\\
	&w_0\to w_0J_0(a_{AA}A)
	\end{aligned},
\end{equation}
where $J_0$ is the zeroth Bessel function of the first kind. Therefore, in the high frequency regime, the interlayer couplings are suppressed by this type of electromagnetic field. This can lead to a renormalization of the bandwidth. In Fig. \ref{fig:wg_qe}, we plot the quasienergies near the Floquet zone center along a high symmetry path in the mBZ. The waveguide drive renormalized the bands, without breaking the symmetries of the static system. The renormalization of the bandwidth depends on the twist angle. In Fig. \ref{fig:wg_bw}, we show the bandwidth of the four bands closest to the Floquet zone center as a function of the twist angle between the layers considering the effect of trigonal warping and particle-hole symmetry breaking terms. These results suggest that dynamical bandwidth tuning could be achieved in TDBG samples without breaking the symmetries of the static system.

\section{Experimental parameter estimates}
 \label{sec:exp}

The laser drive parameters required to obtain the effects discussed here are accessible in experimental settings. The quasienergy gap closings for TDBG without applied static electric field where obtained for driving strengths $(e/\hbar) a_0 A =e a_0 E/( \hbar \Omega) \lesssim 0.06$. For the high frequency considered in the UV regime, $\hbar \Omega = 2 W \approx 5350$~meV, a peak electric field $E\approx 9.5$MV/cm leads to the required driving strength. A combination of a stronger electric field and lower driving frequency could be also considered. For example, to obtain Floquet flat bands, as shown in Fig. \ref{fig:qe_inter}, we use $\hbar \Omega = 1.05 W \approx 2809$~meV and driving strength $(e/\hbar) a_0 A =e a_0 E/( \hbar \Omega) \approx 0.3$ which cab be obtain with a peak electric field $E\approx 25$MV/cm. In graphene, laser pulses with peak electric fields of 30 MV/cm with near-IR frequencies have been employed to generate light-field-driven currents~\cite{Higuchi2017}.

\section{Conclusions}
 \label{sec:conclusion}
 
In conclusion, we have studied twisted double bilayer graphene driven by circularly polarized light in free space and confined into a waveguide. For light propagating in free space, we demonstrated that TDBG in the AB/AB configuration with an applied static electric field perpendicular to the layers, the drive permits  valley-selective quasienergy gap engineering. The periodic drive also leads to a topological transition into a Chern insulating state due to the broken time-reversal symmetry. For TDBG in the AB/BA configuration, the driving protocol can lead to valley-selective engineering even in the absence of an applied electric field. Finally, we showed that stronger drives can generate Floquet flat bands. On the other hand, light confined into a waveguide allows us to dynamically tune the bandwidth of the Floquet zone center bands without breaking the symmetries of the static system. Therefore, employing two complementary Floquet protocols, we showed that light-driven TDBG is a flexible system that could be used as a platform to generate valley-polarized currents.

\section{Acknowledgements} 

This research was primarily supported by the National Science Foundation through the Center for Dynamics and Control of Materials: an NSF MRSEC under Cooperative Agreement No. DMR-1720595, with additional support from NSF DMR-1949701.



%

\appendix
\section{Effective Floquet Hamiltonian in the intermediate frequency regime}
\label{app:effective}

For simplicity of discussion we will neglect the effects of trigonal warping for this section. That is the model Hamiltonian we consider is
\begin{equation}
H_{0,s} = 
\begin{pmatrix} 
h^{+}_{t,s}  + \Delta_1  & t_s  &  0   & 0  \\ 
t_s^\dag  & h^+_{b,s} + \Delta_2   &  T({\bf r})   & 0  \\ 
0  &  T^{\dag}({\bf r})  &  h^-_{t,s}  + \Delta_3    &   t_s  \\
0  & 0 &  t_s^\dag   &   h^-_{b,s} + \Delta_4  \\ 
\end{pmatrix},
\label{hamiltonian}
\end{equation}

where $h_{l,s}^d=v_F(R(-d \theta/2)\vect (k-\kappa_d))\sigma_{xy}+\delta(\mathbb{1}- l s\sigma_3)$ is the Hamiltonian for graphene and the index $d=\pm$ labels the two bilayer graphenes and  $l=\pm$ (with "$+$" for t and "$-$" for b) labels the layers of each double layer.  Furthermore we distinguish between AB and BA stacking for the double layers via a term $s$ that is $s=\pm$ for AB/BA stacking. Therefore $\delta l s\sigma_3$ is a stacking and layer dependent gap, $t$ is the interlayer hopping matrix for the top and bottom double layers with $t_s=t/2(\sigma_1-i s\sigma_2)$. The term 
 \begin{equation}
 	\begin{aligned}
 		&T({\bf r})=\sum_{i=-1}^1 e^{-i \vect { Q}_i \vect r}T_i\\
 		&T_n=\omega^\prime\mathbb{1}_2+\omega\left[\cos\left(\frac{2\pi n}{3}\right)\sigma_1+\sin\left(\frac{2\pi n}{3}\right)\sigma_2\right]
 	\end{aligned}
 \end{equation}
  describes the hopping between the two bilayers and captures the spatial dependence due to the mutual rotation. It depends on two coupling strengths $\omega$ and $\omega^\prime$ that capture the effect that AB/BA and AA-type regions of the center twisted bilayer can have different lattice constants. Since the hopping is dominated by hopping between adjacent layers we neglect higher order interlayer tunnelings. Lastly $\Delta_i$ describe a layer dependent bias.
 
 We introduce circularly polarized light by means of minimal substitution $\vect p\to \vect p+\vect A/v_F$ with $\vect A=A(\cos(\Omega t),\sin(\Omega))$. The Hamiltonian then becomes periodically time dependent $H(t)=H(t+T)$ with period $T$ and can be split as $H(t)=H_0+V(t)$, where
 \begin{equation}
 	V(t)=\begin{pmatrix}
 	v^+(t)&0&0&0\\
 	0&v^+(t)&0&0\\
 	0&0&v^-(t)&0\\
 	0&0&0&v^-(t)
 	\end{pmatrix},
 \end{equation}
 where $v^\pm(t)=R(\mp \theta)\vect A\sigma_{xy}$. 
 
 The time dependence makes a full treatment of the problem cumbersome especially if one wants to build on the model and introduce additional complications such as disorder. Luckily the time dependence can be reduced. This can either be done via a perturbative expansion or non-perturbatively by going to a rotating frame. A useful rotating frame of such a sort is implemented by a unitary transformation $U(t)$ that fulfils $U(T)=\mathbb{1}$ because at stroboscopic times one may forget about the unitary transformation since it is unity. Naively one may choose such a unitary transform that has the form $U(t)=e^{-i\int dt V(t)}$. While this is useful and leads to good results it is not the ideal choice for the problem at hand. This is because it can introduce mathematical artifacts such as an unphysical breaking of rotational symmetry like in the case of graphene [cite our paper and eckhardt's]. For our case we therefore employ a better choice that was introduced in [cite our paper] and split the time dependent part of the Hamiltonian as
 \begin{equation}
 	\begin{aligned}
 	&V_1(t)=A\cos(\omega t)\begin{pmatrix}
 	\sigma_1^{\theta/2}&0&0&0\\
 	0&\sigma_1^{\theta/2}&0&0\\
 	0&0&\sigma_1^{-\theta/2}&0\\
 	0&0&0&\sigma_1^{-\theta/2}
 	\end{pmatrix}\\
 	&V_2(t)=A\sin(\omega t)\begin{pmatrix}
 	\sigma_2^{\theta/2}&0&0&0\\
 	0&\sigma_2^{\theta/2}&0&0\\
 	0&0&\sigma_2^{-\theta/2}&0\\
 	0&0&0&\sigma_2^{-\theta/2}
 	\end{pmatrix}
 	\end{aligned},
 \end{equation}
 
 where the rotated Pauli matrices $\sigma_i^{\theta/2}$ given by $\sigma_i^{\theta/2}=e^{i/4\theta\sigma_3}\sigma_i e^{-i/4\theta\sigma_3}$ were introduced as a convenient shorthand. The unitary transformation we use now is given as
 \begin{equation}
 	U(t)=e^{-i\int dt V_1(t)}e^{-i\int dt V_2(t)}.
 \end{equation}
This choice is useful because it preserves rotational invariance of the dispersion relation after a time average over one period if the interlayer couplings are neglected as seen in [cite our paper] and leads to an improvement over more conventional high frequency expansions. 

If we apply this unitary transformation to the Schr\"odinger equation $i\partial_t \psi=(H_0+V(t))\psi$ and take an average over one period we arrive at the following effective Hamiltonian
\begin{equation}
	H_{F,s} = 
	R^\dag \begin{pmatrix} 
	\tilde h^+_{t,s}  + \Delta_1  & \tilde t_s  &  0   & 0  \\ 
	\tilde t_s^\dag  & \tilde h^+_{b,s} + \Delta_2   &  \tilde T({\bf r})   & 0  \\ 
	0  &  \tilde T^{\dag}({\bf r})  &  \tilde h^-_{t,s}  + \Delta_3    &   \tilde t_s  \\
	0  & 0 &  \tilde t_s^\dag   &   \tilde h^-_{b,s} + \Delta_4  \\ 
	\end{pmatrix}R,
\end{equation}

where $R$ is a unitary transformation given as
\begin{equation}
	R=\begin{pmatrix}
	R_+&0&0&0\\
	0&R_+&0&0\\
	0&0&R_-&0\\
	0&0&0&R_-
	\end{pmatrix};\quad R_\pm=\exp(-\frac{i}{2}\sigma_2^{\pm\theta/2}\tilde A)
\end{equation}

which includes a rotation around the y-axis in pseudospin space with angle $\tilde A=\frac{2A}{\Omega}$. In this rotated space we find that the Hamiltonian for a single graphene layer is modified as
\begin{equation}
	\tilde h_{l,s}^d=\tilde v_F(R(-d\;\theta/2)\vect k)\sigma_{xy}+\delta(\mathbb{1}- l sJ_0(\sqrt{2}\tilde A)\sigma_3)-\tilde \Delta \sigma_3,
\end{equation}

where we find that the layer and stacking dependent gap $l s\sigma_3\to l sJ_0(\sqrt{2}\tilde A)$ has been suppressed by $J_0(\sqrt{2}\tilde A)$, where $J_i$ are Bessel functions of the first kind. The Fermi velocity is lowered to $v_F\to\tilde v_F=J_0(\tilde A)v_F$ and a new stacking and layer independent gap $\tilde \Delta=A\frac{J_1\left(\frac{2 \sqrt{2} A}{\Omega }\right)}{\sqrt{2}}$ has been introduced. For the interlayer couplings in the two double layers we find $t_s=\tilde t/2(\sigma_1-i s\sigma_2)$ with merely the strength renormalized to $t\to \tilde t=tJ_0(\tilde A)$. The position dependent interlayer coupling for the two center grephene layers changes to
\begin{equation}
\begin{aligned}
&\tilde T({\bf r})=\sum_{i=-1}^1 e^{-i \vect { Q}_i \vect r}(\tilde T_i+\omega_{\theta/2}^{\prime\prime}(\sin(\theta/2)\mathbb{1}_2-i\cos(\theta/2)\sigma_3)\\
&\tilde T_n=\omega^\prime\mathbb{1}_2+\tilde \omega\left[\cos\left(\frac{2\pi n}{3}\right)\sigma_1+\sin\left(\frac{2\pi n}{3}\right)\sigma_2\right]
\end{aligned},
\end{equation}

where the coupling $\omega\to \tilde\omega=J_0(\tilde A)\omega$ has been renormalized and a new angle dependent coupling $\omega^{\prime\prime}_{\theta/2}=\sin (\theta/2 ) \left(J_0\left(\sqrt{2}\tilde A\right)-1\right)\omega^\prime$ has been introduced.

\end{document}